\documentclass[aps,pre,a4paper,twocolumn,twoside,
superscriptaddress,showkeys,floatfix]{revtex4}
\usepackage{amsmath,amssymb}
\usepackage{dcolumn}
\usepackage{graphicx}
\newcommand*{\be}{\begin{equation}}
\newcommand*{\ee}{\end{equation}}
\newcommand*{\ba}{\begin{array}}
\newcommand*{\ea}{\end{array}}
\newcommand*{\bea}{\begin{eqnarray}}
\newcommand*{\eea}{\end{eqnarray}}
\newcommand*{\bean}{\begin{eqnarray*}}
\newcommand*{\eean}{\end{eqnarray*}}

\newcommand*{\lp}{\left(}
\newcommand*{\rp}{\right)}
\newcommand*{\ls}{\left[}
\newcommand*{\rs}{\right]}

\newcommand*{\la}{\langle}
\newcommand*{\La}{\left\la}
\newcommand*{\ra}{\rangle}
\newcommand*{\Ra}{\right\ra}

\def\tens#1{\mbox{\usefont{T1}{cmss}{bx}{n}#1}}

\renewcommand*{\d}{\textrm{d}}
\newcommand*{\e}{\mathrm{e}}
\newcommand*{\veps}{\varepsilon}

\newlength{\glength}
\newcommand*{\CPL}{Chem. Phys. Lett.\ }

\newcommand*{\JCP}{J. Chem. Phys.\ }
\newcommand*{\JML}{J. Mol. Liquids\ }

\newcommand*{\JPC}{J. Phys. Chem.\ }
\newcommand*{\JPCA}{J. Phys. Chem. A\ }
\newcommand*{\JPCB}{J. Phys. Chem. B\ }
\newcommand*{\MP}{Mol. Phys.\ }

\newcommand*{\PRA}{Phys. Rev. A\ }
\newcommand*{\PRE}{Phys. Rev. E\ }

\newcommand*{\PTP}{Progr. Theor. Phys.\ }

\begin{document}
\title{Study of anomalous mobility of polar molecular solutions\\
by means of the site-site memory equation formalism}

\author{A. E. Kobryn}
\affiliation{Department of Theoretical Study, Institute for Molecular
Science, Myodaiji, Okazaki, Aichi 444-8585, Japan}

\author{T. Yamaguchi}
\affiliation{Department of Molecular Design and Engineering,
Graduate School of Engineering,\\
Nagoya University, Chikusa, Nagoya, Aichi 464-8603, Japan}

\author{F. Hirata}
\thanks{The author to whom correspondence should be sent}
\email[\newline Electronic address: ]{hirata@ims.ac.jp}
\homepage{daisy.ims.ac.jp}

\affiliation{Department of Theoretical Study, Institute for Molecular
Science, Myodaiji, Okazaki, Aichi 444-8585, Japan}

\date{November 11, 2004}

\begin{abstract}
In this work, the memory equation approach is applied for theoretical
study of dynamics of polar molecular liquids described by the interaction
site model. The study includes the
tem\-pe\-ra\-tu\-re-den\-si\-ty(pres\-su\-re) dependence of the
translational diffusion coefficients $D$ and orientational relaxation
times $\tau$ for infinitely dilute solutions of acetonitrile and methanol
in water, and methanol in acetonitrile. Calculations are performed over
the range of temperatures and densities employing the SPC/E model for
water and optimized site-site potentials for acetonitrile and methanol.
Despite an approximate cha\-rac\-ter of the model potentials and closure
relation used, the theory is able to reproduce qualitatively all main
features of temperature and density dependences of $D$ and $\tau$
observed in computer and real experiments. In particular, anomalous
behavior, i.e. the increase in mobility with density(pressure), is
obtained for $D$ and $\tau$ of methanol in water, while acetonitrile in
water or methanol in acetonitrile do not show deviations from the usual.
The observed enhancement in the molecular mobility is interpreted in
accordance with the concept by Yamaguchi \textit{et~al.} [J. Chem. Phys.
{119} (2003) 1021], i.e. in terms of two competing origins of friction,
which interplay with each other as density increases: the collisional and
dielectric frictions that have tendency, respectively, to strengthen and
weaken with increasing density.
\end{abstract}



\maketitle

\section{Introduction}
\label{Section01}

The enhancement of the molecular mobility of polar molecules in dense
dipolar liquids under pressure, called anomalous molecular mobility, has
recently attracted considerable attention in solution chemistry
experiments and simulations
\cite{Easteal85,Wakai94,Wakai97,Wakai99,Harris97,Harris98,Harris99a,Harris99b,Chowdhuri03}.
Conventional continuum-based theories of liquid matter collapsed when
applied for the description of anomalous molecular behavior, so that to
make theoretical justification of the observed processes is topical from
the point of view stationed on the statistical mechanical approach
\cite{Hirata03}. In order to clarify the microscopic structure of water,
which is a typical representative of a liquid with anomalous behavior,
and its mobility, it has been studied recently by Yamaguchi
\textit{et~al.} using the memory equation approach for molecular liquids
described by the interaction site model \cite{Yamaguchi03a}. The authors
concluded, that anomalous pressure dependence of dynamic properties of
water can be explained by the strong interatomic Coulomb interaction
correlated with density fluctuations. In particular, when the
number-density fluctuation in the low wave-number region decreases with
increasing pressure, the Coulomb coupling between the low wave-number
charge-density fluctuation and hydrodynamic dielectric mode is reduced.
In the case of another polar liquids, like acetonitrile, the effect of
Coulomb interaction can be blocked by even more substantial repulsive
interaction, which makes it dissimilar with water. Phrasing it
differently -- anomalous molecular mobility strongly depends on the
molecular geometry and dominant type of interatomic interaction, and more
studies is required for better understanding of its nature. In this paper
we pay attention to the study of anomalous pressure behavior of polar
binary molecular solutions by considering such popular models as methanol
in water, acetonitrile in water and methanol in acetonitrile. In
particular, we are interested in the den\-si\-ty(pres\-su\-re) dependence
of translational diffusion coefficients and reorientation relaxation
times of solutes. The rest of the paper is organized as follows. In
section \ref{Section02} we write down equations of motion for site-site
intermediate scattering functions, and give relations for translational
diffusion coefficient and reorientation correlation time in terms of
these functions. Equilibrium properties of the system are obtain using
DRISM, and memory kernels are esti\-ma\-ted by the mode-coupling
approximation. Set up for systems under consideration and details of
numerical procedures are described in section \ref{Section03}. Obtained
results are discus\-sed in section \ref{Section04}, and
conclusions are given in section~\ref{Section05}.

\section{Theory}
\label{Section02}


For the description of time-dependent properties of the system, we will
follow the formalism by Hirata and Chong \cite{Chong03}, which is the
unification of the theory of dynamical processes in simple liquids based
on the Mori approach \cite{Mori65a,Mori65b} and reference interaction
site model for molecular liquids \cite{Hirata81,Hirata82,Hirata83}. In
their formalism, parameters of shortened description of the system are
partial number densities and longitudinal current densities. Then, of
practical interest are elements of the matrix of the site-site
intermediate scattering functions $\tens{F}(k;t)$ and the matrix of their
self-parts $\tens{F}_{\mathrm{s}}(k;t)$ defined by
\begin{subequations}
\begin{eqnarray}
F^{\alpha\gamma}(k;t)&=&
\frac1N\la\rho^{\alpha,*}(\textbf{k};0)\rho^\gamma(\textbf{k};t)\ra,\\
F^{\alpha\gamma}_{\mathrm{s}}(k;t)&=&
\frac1N\la\rho^{\alpha,*}(\textbf{k};0)\rho^\gamma(\textbf{k};t)\ra_{\mathrm{s}},
\end{eqnarray}
\end{subequations}
where $N$ is the total number of particles, $\rho^\alpha(\textbf{k};t)$
is the site $\alpha$ number density in reciprocal space, $\textbf{k}$ is
the wave-vector, $k=|\textbf{k}|$, $t$ is time, $*$ means complex
conjugation, angular brackets denote appropriate statistical ave\-ra\-ge
(e.g., canonical Gibbs ensemble average), and suffix ``s'' stands for
``self'' and means correlations between two sites of a same molecule. In
the limit of infinite dilution, which is considered here, the solution
density is essentially determined by the solvent. Consecutively, one has
two types of memory equations for the solvent subsystem and one for the
solute subsystem (indicated by the superscript ``u'') which are:
\begin{subequations}
\label{memeq}
\begin{align}
\ddot{\tens{F}}(k;t)&=-\la\boldsymbol{\omega}_k^2\ra\tens{F}(k;t)
-\int_0^t\d\tau\;\tens{K}(k;\tau)\dot{\tens{F}}(k;t-\tau),\nonumber\\\label{mevc}\\
\ddot{\tens{F}}_{\mathrm{s}}(k;t)&=
-\la\boldsymbol{\omega}_k^2\ra_{\mathrm{s}}\tens{F}_{\mathrm{s}}(k;t)
-\int_0^t\d\tau\;\tens{K}_{\mathrm{s}}(k;\tau)\dot{\tens{F}}_{\mathrm{s}}(k;t-\tau),
\nonumber\\\label{mevs}\\
\ddot{\tens{F}}^{\mathrm{u}}_{\mathrm{s}}(k;t)&=
-\la\boldsymbol{\omega}_k^2\ra^{\mathrm{u}}_{\mathrm{s}}\tens{F}^{\mathrm{u}}_{\mathrm{s}}(k;t)
-\int_0^t\d\tau\;\tens{K}^{\mathrm{u}}_{\mathrm{s}}(k;\tau)
\dot{\tens{F}}^{\mathrm{u}}_{\mathrm{s}}(k;t-\tau).\nonumber\\\label{meus}
\end{align}
\end{subequations}
In these equations, dot over the quantity means its time derivative; the
memory function matrices, denoted as $\tens{K}(k;t)$,
$\tens{K}_{\mathrm{s}}(k;t)$ and
$\tens{K}^{\mathrm{u}}_{\mathrm{s}}(k;t)$, describe the friction on the
motion of interaction sites; quantities $\la\boldsymbol{\omega}_k^2\ra$,
$\la\boldsymbol{\omega}_k^2\ra_{\mathrm{s}}$ and
$\la\boldsymbol{\omega}_k^2\ra_{\mathrm{s}}^{\mathrm{u}}$ are normalized
second order frequency matrices given by
\begin{subequations}
\bea
\la\boldsymbol{\omega}_k^2\ra&=&k^2\tens{J}(k)\tens{S}^{-1}(k),\\
\la\boldsymbol{\omega}_k^2\ra_{\mathrm{s}}&=&k^2\tens{J}(k)\tens{S}_\mathrm{s}^{-1}(k),\\
\la\boldsymbol{\omega}_k^2\ra_{\mathrm{s}}^{\mathrm{u}}&=&
k^2\tens{J}^{\mathrm{u}}(k)\tens{S}_\mathrm{s}^{\mathrm{u},-1}(k),
\eea
\end{subequations}
where $\tens{S}(k)\equiv\tens{F}(k;t=0)$,
$\tens{S}_{\mathrm{s}}(k)\equiv\tens{F}_{\mathrm{s}}(k;t=0)$ and
$\tens{S}_{\mathrm{s}}^{\mathrm{u}}(k)\equiv\tens{F}_{\mathrm{s}}^{\mathrm{u}}(k;t=0)$
are matrices of static site-site structure factors and their self parts,
respectively, while $\tens{J}(k)$ is the matrix of static longitudinal
site-current correlation functions
\bea
\lfloor\tens{J}(k)\rfloor^{\alpha\gamma}=
\frac1N\sum_{i,j}\La v^{\alpha}_{i,z}v^{\gamma}_{j,z}
\e^{-i\textbf{k}\cdot(\textbf{r}_i^\alpha-\textbf{r}_j^\gamma)}\Ra,
\label{longitudinal-current}
\eea
where subscripts $i,j$ refer to molecules,
$\textbf{r}_i^\alpha\equiv\textbf{r}_i^\alpha(0)$ is the initial position
of site $\alpha$, and $v^{\alpha}_{i,z}\equiv{}v^{\alpha}_{i,z}(0)$ is
longitudinal component of the initial velocity of site $\alpha$. The
analytical expression of $\tens{J}(k)$ for arbitrary shape of the
molecule has been given recently by Yamaguchi \textit{et al.}
\cite{Yamaguchi04b}. Definition for $\tens{J}^{\mathrm{u}}(k)$ is similar
to the one given by equation (\ref{longitudinal-current}) with the
difference that summation runs over the solute molecule only. Initial
values of intermediate scattering functions, that we need to solve memory
equations (\ref{memeq}), can be obtained using RISM theory
\cite{Hirata81,Hirata82,Hirata83}. That theory predicts static structure
of molecular fluids via the calculation of their site-site pair
correlation functions. In this work, however, in order to have a better
description for the dielectric susceptibility of solvent we employ the
formalism of DRISM that uses experimental value of dielectric constant
\cite{Perkyns92a,Perkyns92b}.

Quantities of our interest in this paper are solute's translational
diffusion coefficient $D$ and rank-1 reorientation correlation time
$\tau$. Following the derivation procedure presented in \cite{Chong98c},
the translational diffusion coefficient $D$ is obtained as
\bea
D&=&\frac13\int_0^\infty\d t\;Z^{\alpha\gamma}(t)\nonumber\\
&=&-\lim_{t\to\infty}\int_0^t\d\tau\lim_{k\to0}\frac1{k^2}
\lfloor\ddot{\tens{F}}_{\mathrm{s}}^{\mathrm{u}}(k;\tau)\rfloor^{\alpha\gamma},
\eea
where $Z^{\alpha\gamma}(t)$ is the site-site velocity autocorrelation
function with sites $\alpha$ and $\gamma$ belonging to the same molecule.
The rank-1 reorientation correlation function $C_{\boldsymbol{\mu}}(t)$
is defined by
\be
C_{\boldsymbol{\mu}}(t)=
\frac{\sum_i\La\boldsymbol{\mu}_i(0)\boldsymbol{\mu}_i(t)\Ra}
{\sum_j\La|\boldsymbol{\mu}_j|^2\Ra},
\label{Cmu}
\ee
where $\boldsymbol{\mu}_i(t)$ is a vector fixed on the molecule $i$. In
our case it is the dipole moment and therefore can be described by the
linear combination of site coordinates as
$\boldsymbol{\mu}_i(t)=\sum_{\alpha}z_\alpha\textbf{r}^\alpha(t)$ with
$z_\alpha$ being site partial charges. Putting that into equation
(\ref{Cmu}) and using properties of time-correlation functions
\cite{Harp70} one arrives at
\begin{subequations}
\bea
C_{\boldsymbol{\mu}}(t)&=&
\frac{\sum_i\sum_{\alpha\gamma}z_{\alpha}z_{\gamma}
\La\textbf{r}_i^\alpha(0)\textbf{r}_i^\gamma(t)\Ra}
{\sum_j\La|\boldsymbol{\mu}_j|^2\Ra},\\
\ddot{C}_{\boldsymbol{\mu}}(t)&=&
-\frac{N\sum_{\alpha\gamma}z_{\alpha}z_{\gamma}Z^{\alpha\gamma}(t)}
{\sum_j\La|\boldsymbol{\mu}_j|^2\Ra}.
\eea
\end{subequations}
Hence, the time development of both $Z^{\alpha\gamma}(t)$ and
$C_{\boldsymbol{\mu}}(t)$ is governed by the memory equation for the
self-part of the site-site intermediate scattering function
$\tens{F}_{\mathrm{s}}^{\mathrm{u}}(k;t)$. The rank-1 reorientation
relaxation time is defined as \cite{Hansen86}
\be
\tau=\int_0^\infty\d t\;C_{\boldsymbol{\mu}}(t).
\ee

Since the dynamics we are interested in corresponds to the long-time
limit, memory kernels for memory equations can be constructed using the
mode-coupling approximation \cite{Hansen86}. In works by Chong
\textit{et~al.} \cite{Chong98b,Chong02} the conventional mode-coupling
theory has been extended to the case of molecular liquids based on the
interaction-site model. It has been shown, however, that the proposed
expressions for memory functions underestimate friction in orientational
motions \cite{Yamaguchi02b}. In our study we use the recipe by Yamaguchi
and Hirata \cite{Yamaguchi02b}, who suggested a modified expression that
includes the interaxial coupling.

\section{Details of the model setup and numerical procedures}
\label{Section03}

We performed calculations for acetonitrile in water, methanol in water
and methanol in acetonitrile, all in the case of infinite dilution. As
for the structure and the intermolecular potential of water we employed a
model of the extended simple point charge (SPC/E) \cite{Berendsen87}. We
also put the Lennard-Jones (LJ) core on the hydrogen atoms in order to
avoid the undesired divergence of the solution of the DRISM integral
equation. The LJ parameters of the hydrogen atom, the depth of the well
and the diameter, were chosen to be $0.046$ kcal/mol and $0.7$ \AA,
respectively.

In acetonitrile and methanol the methyl group was considered to be a
single interaction site located on the methyl carbon atom. So that all
chemical compounds consisted of three sites which interact through the
pair potential \cite{Edwards84,Jorgensen86}
\begin{equation}
\phi(r_{ij}^{\alpha\gamma})=
4\epsilon_{\alpha\gamma}\ls\lp\frac{\sigma_{\alpha\gamma}}{r_{ij}^{\alpha\gamma}}\rp^{12}
-\lp\frac{\sigma_{\alpha\gamma}}{r_{ij}^{\alpha\gamma}}\rp^{6}\rs
+\frac{z_{\alpha}z_{\gamma}}{r_{ij}^{\alpha\gamma}},
\label{interaction-potential}
\end{equation}
i.e., LJ plus Coulomb. Here
$r_{ij}^{\alpha\gamma}=|\mathbf{r}_i^\alpha-\mathbf{r}_j^\gamma|$;
parameters $\epsilon_{\alpha\gamma}$ and $\sigma_{\alpha\gamma}$ are LJ
well-depths and LJ diameters defined as
$\epsilon_{\alpha\gamma}=\sqrt{\epsilon_\alpha\epsilon_\gamma}$ and
$\sigma_{\alpha\gamma}=(\sigma_\alpha+\sigma_\gamma)/2$, respectively,
with $\sigma_\alpha$ being the LJ diameter of a single site. Point
charges for acetonitrile were chosen to reproduce electrostatic potential
obtained in {\itshape ab initio} calculations \cite{Edwards84}.

In calculations for acetonitrile or methanol in water temperature of the
system was varied from $258.15$ to $373.15$ K, and density of water from
$0.9$ to $1.2$ g/cm${}^3$; for the case of methanol in acetonitrile
temperature of the system was varied from $293.15$ to $323.15$ K, and
density of acetonitrile from $0.6726$ to $0.815$ g/cm${}^3$. Connection
of the water parameters with thermodynamic pressure can be established,
e.g., using the multi-parametric equation of state \cite{Wagner02}
(except for the metastable regions where reliable data is lacking).

Tem\-pe\-ra\-tu\-re/den\-si\-ty dependent dielectric constant
$\varepsilon$ for water used in numerical calculations has been evaluated
as a physical solution of an empirical nonlinear equation
\cite{LandoltBornstein80}:
\begin{equation}
\veps-\frac12\lp1+\frac1{\veps}\rp=\frac1v
\lp17+\frac{9.32\cdot10^4\lp1+\frac{153}{v\cdot{T}^{1/4}}\rp}{\lp1-3/v\rp^2T}\rp,
\end{equation}
where $v$ is a molar volume in units of cm$^3$/mol, and $T$ is
thermodynamic temperature~in~K. This equation has been used also in such
tem\-pe\-ra\-tu\-re/den\-si\-ty points where no experimental values
exist. For density and dielectric constant of acetonitrile at different
temperatures we used experimental values indicated in
Table~\ref{density-epsilon-acetonitrile}.
\begin{table}[!htb]
\fontsize{8}{9.6}\selectfont
\caption{Density and dielectric constant for acetonitrile as functions of temperature:
experimental data used in our computation. Temperature is in K, and density is in g/cm$^3$.}
\label{density-epsilon-acetonitrile}
\begin{ruledtabular}
\begin{tabular}{lccccccc}
$T$&\;\;293.13&\;\;295.05&\;\;298.15&\;\;303.15&\;\;308.15&\;\;313.15&\;\;323.15\\\hline
$\rho$&\;.782$^{\mathrm{a}}$&---&\;.7762$^{\mathrm{b}}$&\;.7712$^{\mathrm{a}}$&\;.7652$^{\mathrm{b}}$&\;.7603$^{\mathrm{a}}$&\;.7492$^{\mathrm{a}}$\\
$\veps$&\;38.8$^{\mathrm{c}}$&\;37.5$^{\mathrm{d}}$&\;36.69$^{\mathrm{e}}$&\;35.93$^{\mathrm{e}}$&&&\;33.5$^{\mathrm{f}}$\\
\end{tabular}
\end{ruledtabular}
$^{\mathrm{a}}$\cite{Ku98}, 
$^{\mathrm{b}}$\cite{Nikam98},
$^{\mathrm{c}}$\cite{Delta},
$^{\mathrm{d}}$\cite{ASI},
$^{\mathrm{e}}$\cite{Cavell65},
$^{\mathrm{f}}$Extrapolated\hfill~
\end{table}

In the numerical part of this study we calculated first the site-site
static structure factor by solving the DRISM equation using the
intermolecular interaction, molecular shape, temperature and density. In
order to improve the convergence of the DRISM calculation, we used the
method of the modified direct inversion in an iterative space (MDIIS)
proposed by Kovalenko \textit{et al.} \cite{Kovalenko99}. The value of
the parameter of DRISM theory $a$ has been set to $0.1$ \AA$^2$. From the
static site-site structure factor, we calculated the site-site
intermediate scattering function using the site-site memory equation with
the mode-coupling approximation for the memory kernels. The memory
equation has been time-integrated numerically. Time-development of
correlation functions in the $k\to0$ limit was treated separately by the
analytical limiting procedure of theoretical expressions. In the
numerical procedure, the reciprocal space was linearly discretized as
$k=(n+\frac12)\Delta k$, where $n$ is an integer from 0 to $N_k-1$.
Values of $\Delta k$ and $N_k$ are $0.061$ \AA${}^{-1}$ and $2^9=512$,
respectively. The choice for $N_k$ as the power of two came as the
requirement of the subroutine for the fast Fourier transform, which has
been used in DRISM / MDIIS. The diffusion coefficient $D$ was calculated
from the asymptotic slope of the time dependence of the mean square
displacement, and the orientational relaxation time $\tau$ was obtained
by time-integration of the rotational autocorrelation function.

\section{Results and discussions}
\label{Section04}

\subsection{Structural properties}

\begin{figure*}[!ht]
\begin{center}
\includegraphics*[bb=14 58 582 260,width=0.9\textwidth]{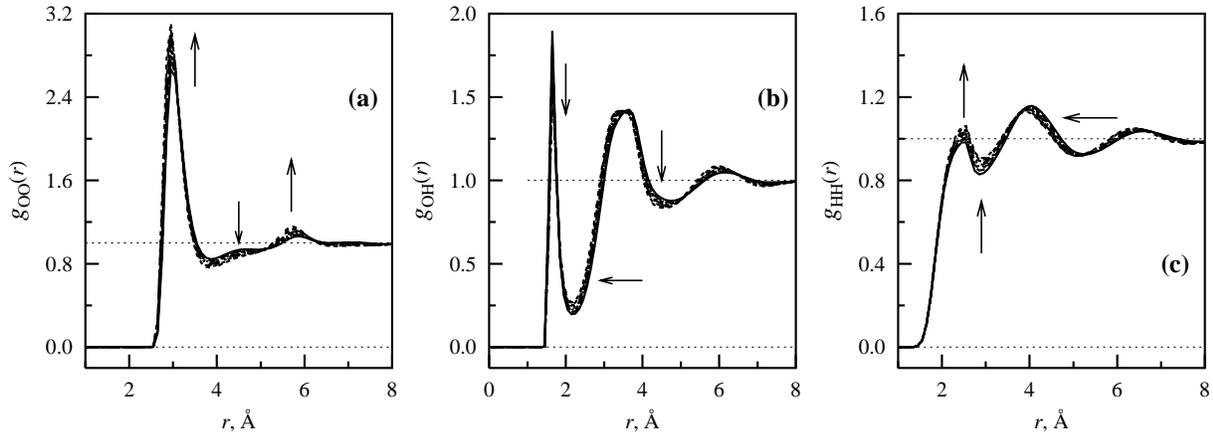}
\end{center}
\caption{Site-site radial distribution functions of neat water at
$T=273.15$ K and set of densities from 0.9 to 1.125 g/cm$^3$, obtained by
the DRISM/HNC integral equation theory. Arrows show directions of
alternations due to an increase in pressure.} \label{gww}
\end{figure*}
 
\begin{figure*}[!ht]
\begin{center}
\includegraphics*[bb=14 58 582 476,width=0.9\textwidth]{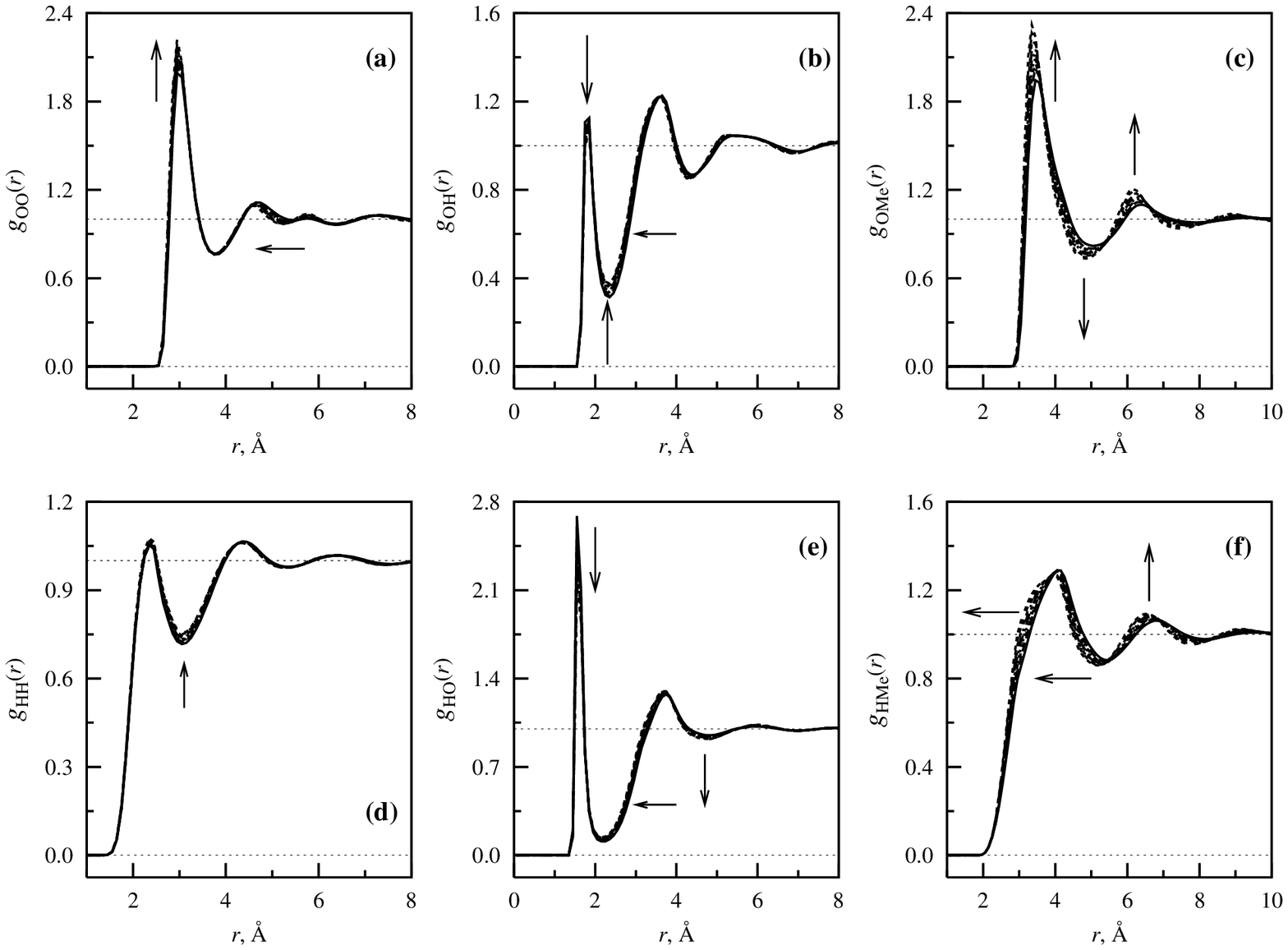}
\end{center}
\caption{Water-methanol site-site radial distribution functions at
$T=273.15$ K and set of densities from 0.9 to 1.125 g/cm$^3$ for water,
obtained by the DRISM/HNC integral equation theory. In the notations
used, first site always belongs to water, and second site always belongs
to methanol. Arrows show directions of al\-ter\-na\-ti\-ons due to an
increase in pressure.} \label{gwm}
\end{figure*}

\begin{figure*}[!ht]
\begin{center}
\includegraphics*[bb=14 58 582 476,width=0.9\textwidth]{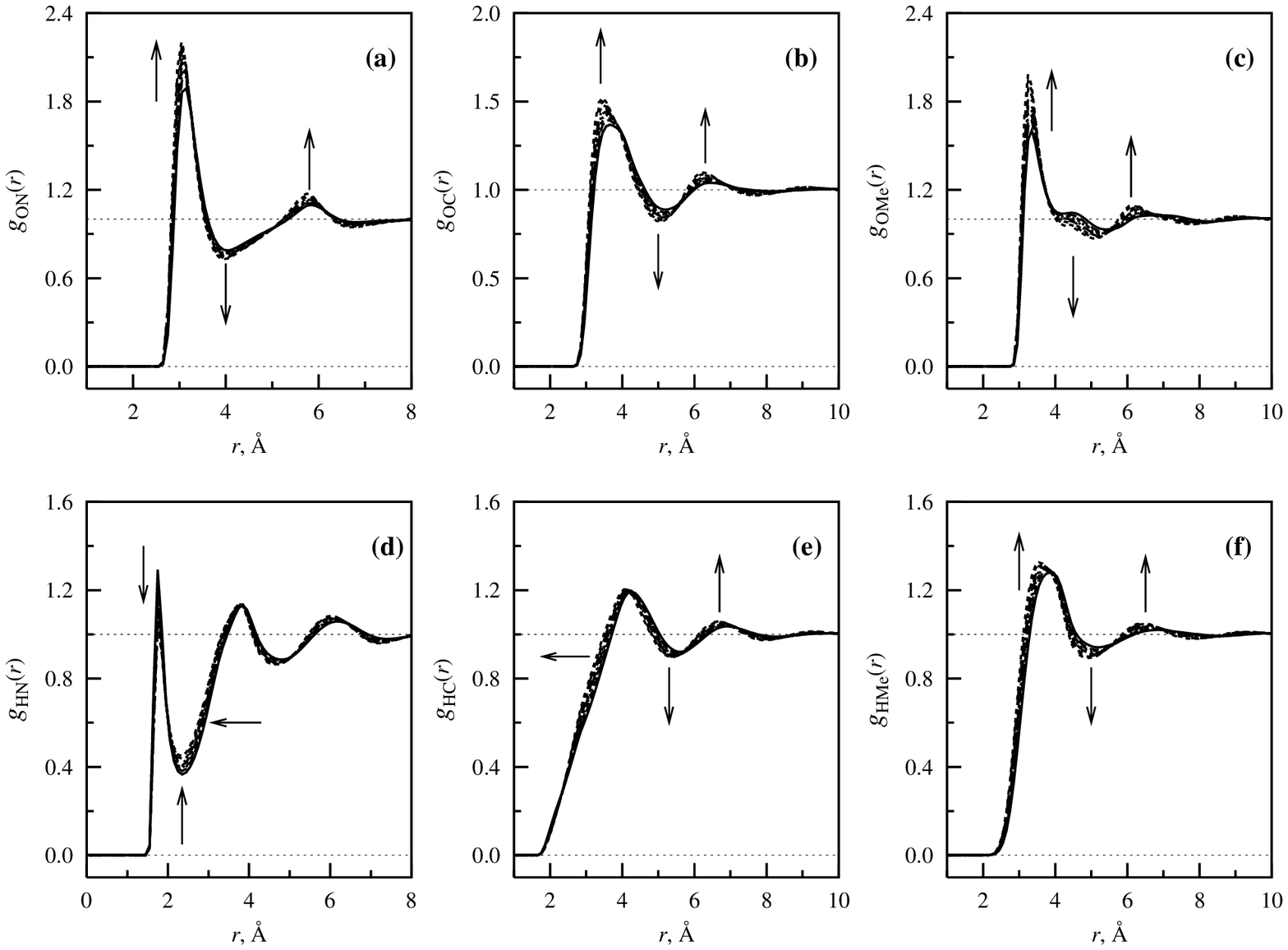}
\end{center}
\caption{Water-acetonitrile site-site radial distribution functions at
$T=273.15$ K and set of densities from 0.9 to 1.125 g/cm$^3$ for water,
obtained by the DRISM/HNC integral equation theory. In the notations
used, first site always belongs to water, and second site always belongs
to acetonitrile. Arrows show directions of al\-ter\-na\-ti\-ons due to an
increase in pressure.} \label{gwa}
\end{figure*}

\begin{figure*}[!ht]
\begin{center}
\includegraphics*[bb=14 58 582 692,width=0.9\textwidth]{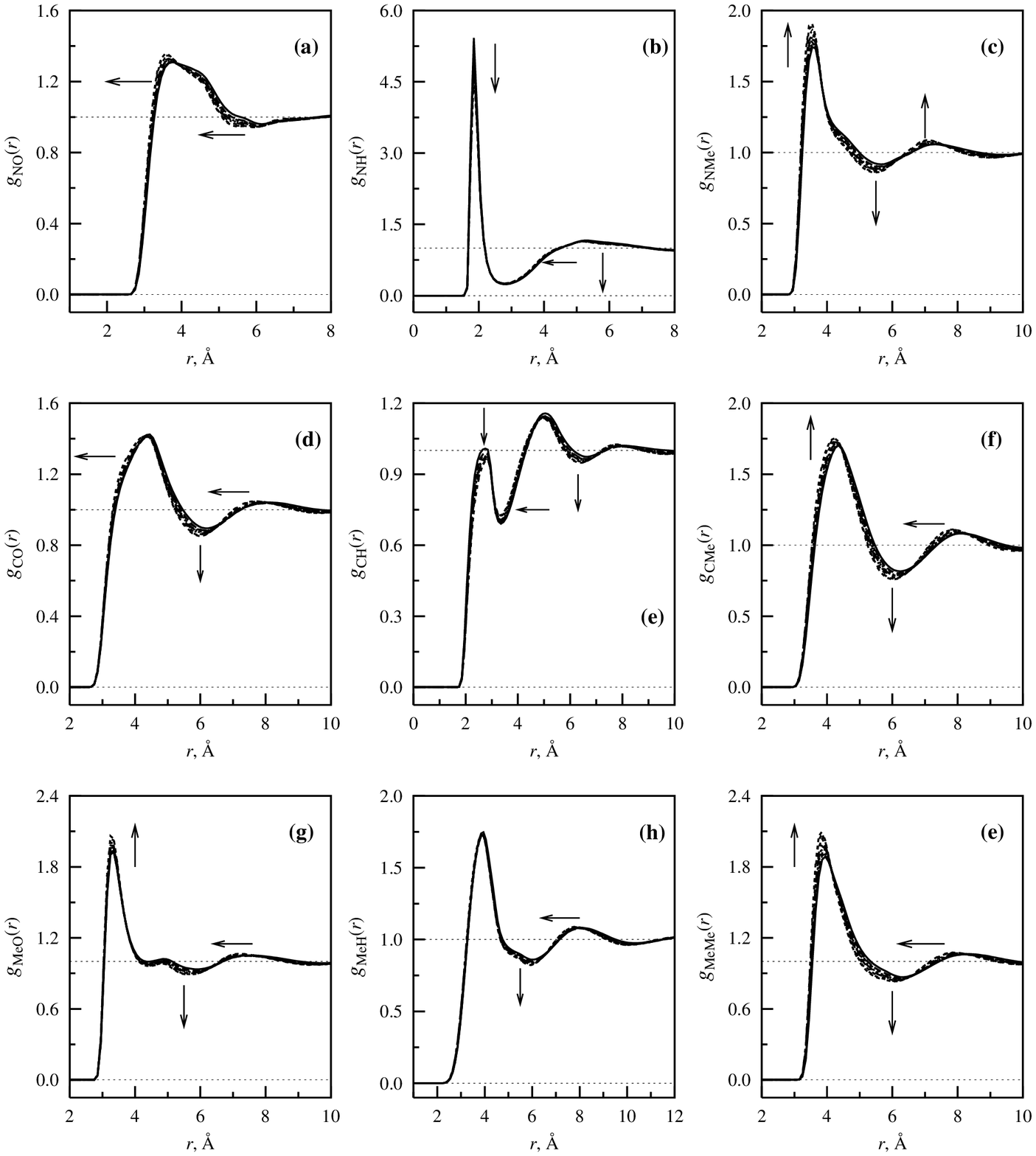}
\end{center}
\caption{Acetonitrile-methanol site-site radial distribution functions at
$T=293.15$ K and set of densities from 0.6726 to 0.815 g/cm$^3$ for
acetonitrile, obtained by the DRISM/HNC 
theory. In the notations used, first site always belongs to acetonitrile,
and second site always belongs to methanol. Arrows show directions of
alternations due to an increase in pressure.} \label{gam}
\end{figure*}

When applied to the liquid state, external pressure induces changes both
in its structure and dynamics of molecules. The \textit{anomalous}
behavior of the molecular mobility with pressure, i.e. dynamics, has been
assigned to the hydrogen bonding properties of the system, which in its
turn can be related to structural properties. The idea is that the
hydrogen bond or hydrogen bonding network are distorted upon compression,
so that the molecule has less hydrogen bonds at higher pressure, which
makes its translational and rotational motion looser. On the other hand,
extensive compression results in significant grow of repulsive forces
between molecules. In fact, the liquid structure is determined
essentially by these two competing factors: hydrogen bonding and
repulsive interaction, which interfere each other through the
intramolecular constraints or molecular geometry. There was much
attention to this topic in scientific literature and we shall not repeat
results of these studies here. Instead of that we can focus our analysis
on some properties related to the subject. Figs. \ref{gww} -- \ref{gam}
show site-site RDF's for neat water, water-methanol, water-acetonitrile
and acetonitrile-methanol infinite dilution solutions, respectively,
calculated by the DRISM/NHC theory for set of densities at constant
temperature. Despite the comparatively big number of site-site
combinations, there are some common features in the behavior of each
individual site-site RDF, that can be categorized as follows. First, one
sees that there are RDF's with the first peak either remained almost
unchanged, or enhanced with the increase of density(pressure). Second,
there are first peaks that are located at positions much shorter, than
other peaks, and that all of them subside with density(pressure) a lot.
The former case is a reflection of packing effect. It is because of the
result of pressure pushes the next nearest neighbor toward the central
molecule and makes the first coordination shell to be outlined more
clearly, which is also confirmed by the drift of majority of peak
locations into direction of shorter separations. The effect of pressure
also distorts the \textit{preferable} mutual orientation of molecules,
which does not favor hydrogen bonds. The latter case in given previously
peak categorization is an example of such state of things. To be more
specific, first peaks in Figs. \ref{gww}(b), \ref{gwm}(b) and
\ref{gwm}(e), \ref{gwa}(d), and \ref{gam}(b) all are evidences of
hydrogen bonds in the system, and all of them subside with the increase
of density.
\enlargethispage{1ex}
There is, however, a difference in their nature. For methanol
in water the hydrogen bonding is realized in two ways: once when the
proton acceptor is solvent, i.e. water, Fig. \ref{gwm}(b); and once when
the proton acceptor is the solute, i.e. methanol, Fig. \ref{gwm}(e). In
the first case the peak amplitude is much lower, than in the second case,
indicating that the first type of hydrogen bonding may be somewhat
weaker. Nevertheless, due to the fact that both solvent and solute can
evident as both proton donor and acceptor, they are able to make either
hydrogen bond network, as in the case of neat water, or hydrogen bond
zigzag chains, as in the case of pure methanol or water-methanol mixture.
For acetonitrile in water the hydrogen bonding is realized only in one
way: when the nitrogen site of acetonitrile serves as the proton
acceptor, Fig. \ref{gwa}(d). The peak amplitude of that hydrogen bond is
also lower compared to neat water, Fig. \ref{gww}(b), or water-methanol,
Fig. \ref{gwm}(e), but general behavior of RDF with density(pressure) is
essentially the same. The lack of ability for acetonitrile to serve as
the proton acceptor \textit{and} donor has far reaching consequences and
results in impossibility to create a hydrogen bond network neither in
pure substance or between water an acetonitrile molecules in their
mixture. Thus, the solution of acetonitrile in water represents situation
when hydrogen bonds are realized between both types of molecules, but
hydrogen bond network is possible to establish between water molecules
exclusively. Finally, for methanol in acetonitrile the only hydrogen bond
available in the system is between the nitrogen site of the solvent and
hydrogen site of the solute, Fig. \ref{gam}(b). Any type of the hydrogen
bond network does not appear in such solution at all.

\subsection{Dynamical properties}

\begin{figure*}[!ht]
\begin{center}
\includegraphics*[bb=18 56 572 584,width=0.9\textwidth]{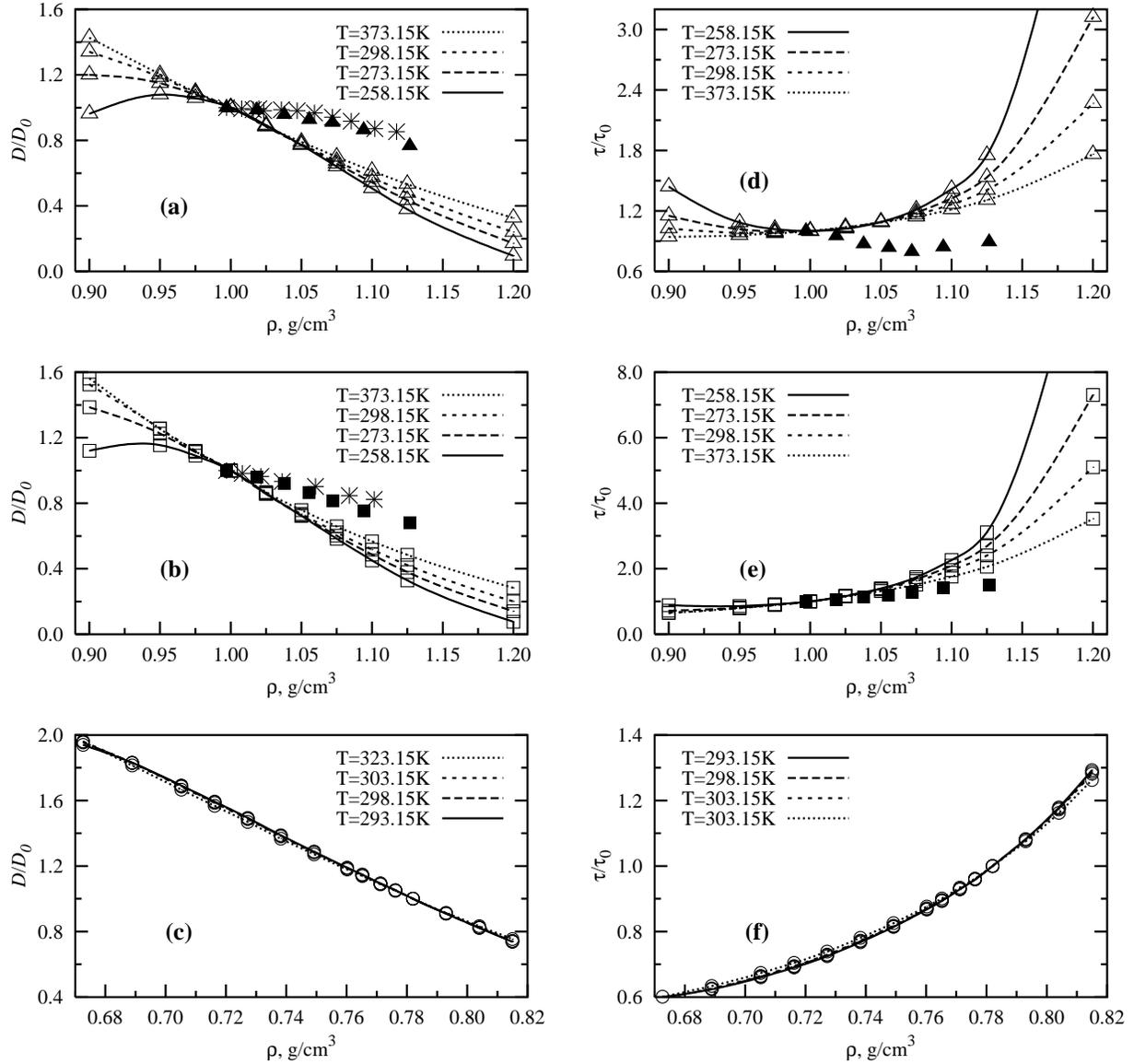}
\end{center}
\caption{Normalized translational diffusion coefficient $D/D_0$ 
and reorientation relaxation time $\tau/\tau_0$ for
for methanol in water (a), (d); acetonitrile in water (b), (e); and for methaol in
acetonitrile (c), (f), respectively. {\footnotesize$\square$}, {\footnotesize$\triangle$},
{\Large$\circ$} -- theory, {\footnotesize$\blacksquare$},
$\blacktriangle$ -- results of MD simulation \protect\cite{Chowdhuri03} for $T=298$ K,
$*$ -- experimental data \protect\cite{Easteal85} for $T=298.2$ K,
lines connecting open symbols are cubic splines for the eye-guide.}
\label{diffusion}
\end{figure*}

Fig. \ref{diffusion} shows the density(pressure) dependence of the
normalized translational diffusion coefficient $D/D_0$ and normalized
reorientation relaxation time $\tau/\tau_0$ for investigated solutes at
various temperatures. Here $D_0$ and $\tau_0$ are translational diffusion
coefficient and reorientation relaxation time, respectively, of the
solute at the ambient density for the solvent, which is $\rho=0.997047$
g/cm$^3$ for water and $\rho=0.782$ g/cm$^3$ for acetonitrile. In
particular, Fig. \ref{diffusion}(a) shows that for sufficiently low
temperature the diffusion coefficient of methanol in water first
increases with density(pressure), and then smoothly changes its behavior
to the normal one, i.e. decreases with density(pressure). Relatively flat
maximum is observed for the density slightly smaller than that at the
ambient condition. At higher temperatures maximum is not observed.
Similar features can be mentioned about behavior of the diffusion
coefficient of acetonitrile in water, Fig. \ref{diffusion}(b). But in
this case the increase of the diffusion coefficient at the lowest
temperature is very small and can be monitored only by comparing numbers
for neighboring density points. The cubic spline curve, which connects
these points, makes this very flat maximum observed visually. Behavior of
the diffusion coefficient at higher temperatures does not deviate from
the normal one, as in the case of methanol in water. Experimental
measurements of diffusion coefficients for acetonitrile in water and
acetonitrile-D$_3$ (CD$_3$CN) in water, both at $T=303$ K, reported by
Nakahara \textit{et~al.} \cite{Wakai94} testify the same tendency in
their density(pressure) behavior as obtained from the theory. Figure
\ref{diffusion}(c) demonstrates monotonous decrease of the diffusion
coefficient of methanol in acetonitrile in the entire range of densities
and at all investigated temperatures. Fig. \ref{diffusion}(d) is a
typical example of anomalous density(pressure) dependence of the
reorientation relaxation time. It is clear to see that it first decreases
with density(pressure) and then starts to behave normally, i.e. increases
with density(pressure). The order of decrease smoothly varies from about
40\% at $T=258.15$ K to about 3\% at $T=298.15$ K. At $T=373.15$ K
anomaly is not observed, but the increase is very slow for the quite wide
range of densities(pressures). In the case of acetonitrile in water, Fig.
\ref{diffusion}(e), normal behavior of the reorientation relaxation time
is observed almost entirely except for the lowest temperature, where
$\tau$ has very shallow minimum. Finally, Fig. \ref{diffusion}(f)
demonstrates monotonous increase of the reorientation relaxation time of
methanol in acetonitrile in the entire range of densities and at all
investigated temperatures. In such a way, it can be regarded as a typical
example of normal density(pressure) be\-ha\-vi\-or of $\tau$. Filled
triangles and filled squares in Fig. \ref{diffusion} are results of MD
simulation \cite{Chowdhuri03} at $T=298$ K for methanol in water and
acetonitrile in water, respectively, both at the infinite dilution. And
asterisk symbols are experimental measurements \cite{Easteal85} at
$T=298.2$ K. One can see rather satisfactory correlation between the
results of our computation, MD simulations and experiment, both
qualitatively and quantitatively. It should be noted, however, that in
terms of absolute values of $D$ and $\tau$ agreement with simulation and
experimental data is poor.

Theoretical aspects of anomalous molecular mobility have been considered
previously by the example of neat water \cite{Yamaguchi03a}. Within the
model employed, the authors proposed that the enhancement of the
molecular mobility by compression can be related to the suppression of
the number-density fluctuations in the low-$k$ region ($k$ is the
wave-number), rather than only the breakdown of the tetrahedral
hydrogen-bonding network structure of water. The reasoning of this
statement coherently follows from the examination of the behavior of
memory functions \cite{Yamaguchi03a} and is in harmony with the heuristic
explanation of the picture \cite{Yamaguchi04a}. Let us imagine a molecule
rotating in a polar media. The rotational motion will induce a relaxation
process of surrounding molecules in order to make themselves aligned to
the electric field produced by the new orientation of the molecule in
concern. The energy dissipation associated with the relaxation process is
an origin of the friction on the rotational motion, or the dielectric
friction. The larger is the charge density fluctuation, the greater is
the dielectric friction. When the liquid is compressed by an external
pressure, the mechanism of electrostatic friction on the dielectric
relaxation is mostly the same except that molecules in the solution are
packed more tightly. Higher packing fraction prevents large number
density fluctuations leading to smaller non-uniformity in the the
polarization density and, as the result, smaller heterogeneity in the
charge density. Then the electrostatic friction on dielectric relaxation
is smaller and means that the dielectric relaxation time becomes shorter.
Such an acceleration enhances the mobility of molecules in solution
through the dielectric friction mechanism. Spherical shape of the
molecule, or equivalently -- spherical shape for the repulsive
short-range interaction, favors anomalous molecular mobility. It is
because the dominant contribution to the rotational and therefore
dielectric friction is defined by the type of the interaction and its
range, which explains the difference in the behavior of nearly spherical
shape molecule of methanol and rod-like shape molecule of acetonitrile.
For methanol the dominant contribution comes from the long-range Coulomb
interaction, while for acetonitrile from the short-range repulsive one.
As the result, methanol in water exhibits more substantial anomaly
compared to acetonitrile in water. The anomalous behavior is attributed
to the strong electrostatic interaction -- the ``hydrogen bond'', among
the solvent molecules and those between solute and solvent. The former
causes the decrease in the dielectric relaxation time with pressure,
while the latter induces the coupling between the dielectric mode of the
solvent and the rotation of the solute. The anomaly is largely suppressed
for acetonitrile in water due to enhanced significance of the repulsive
core in the molecule. Finally, methanol in acetonitrile does not show any
indication of the anomalous density dependence because of strong
collisional friction on the collective reorientation of the solvent in
this case, so that the dielectric relaxation becomes slower with
pressure. Higher density or pressure just magnifies the effect of the
repulsive core upon rotation and because of that -- the collisional
friction.

\section{Summary}
\label{Section05}

In present paper we have calculated the den\-si\-ty dependence of the
translational diffusion coefficients and rank-1 reorientation relaxation
times for acetonitrile and methanol in water, and methanol in
acetonitrile at various temperatures. Calculations have been
per\-for\-med using the site-site memory equation with the
mo\-de-coup\-ling approximation for memory kernels, and the DRISM/HNC for
static properties. For simplicity of theoretical and computational
procedures solutions have been considered in the limit of infinite
dilution. Calculated quantities have been obtained to behave anomalously
with density(pressure) for methanol in water. In particular,
translational diffusion coefficient $D$ may increase for a while with
density at low temperature (supercooled region), and reorientation
relaxation time $\tau$ may decrease with density to form a minimum in the
vicinity of ambient condition density for water in the quite wide range
of temperatures. Similar computations for acetonitrile in water show tiny
anomaly in the behavior of $D$ and $\tau$ only at the lowest (supercooled
region) temperature, while for other regions there is no deviation from
the usual behavior. And for methanol in acetonitrile there is no
deviation in the entire region of investigated densities and
temperatures. This picture is consistent with results of experimental
observation and MD simulation, also quantitative agreement is not as good
as qualitative. The physical origin of the anomalous density dependence
of molecular mobility is interpreted in terms of two competing origins of
friction, which interplay with each other as density increases: the
collisional and dielectric frictions which, respectively, strengthen and
weaken with increasing density. Presented results are first in scientific
literature to realize anomalous molecular mobility of polar solute in
water by means of the statistical mechanical theory.

\end{document}